# Chapter 10

# Energy Deposition and Radiation to Electronics

*A. Bignami[1], F. Broggi[1], M. Brugger[2], F. Cerutti[2][*], L.S. Esposito[2], A. Lechner[2], N.V. Mokhov[3], I.L. Rakhno[3], C. Santini[4], E. Skordis[2] and I.S. Tropin[3]*

[1]INFN-LASA, Milan, Italy
[2]CERN, Accelerator & Technology Sector, Geneva, Switzerland
[3]FNAL, Fermi National Accelerator Laboratory, Batavia, USA
[4]Politecnico of Milan, Italy

## 10    Energy deposition and radiation to electronics

### 10.1   Characterization of the radiation source

Proton–proton inelastic collisions taking place in the LHC inside its four detectors generate a large number of secondary particles with average multiplicities of approximately 100 (120) per single proton–proton interaction with 3.5 (7) TeV beams, but with very substantial fluctuations over different events. Moving away from the interaction point (IP), this multiform population evolves, even before touching the surrounding material, because of the decay of unstable particles (in particular neutral pions decaying into photon pairs). Figure 10-1 illustrates the composition of the debris at 5 mm from the point of a 14 TeV centre of mass collision, featuring a ∼30% increase in the number of particles, due to the aforementioned decays, and a clear prevalence of photons (almost 50%) and charged pions (∼35%).

Most of these particles are intercepted by the detector and its forward region shielding, releasing their energy within the experimental cavern. However, the most energetic particles, emitted at small angles with respect to the beam direction, travel farther in the vacuum and reach the accelerator elements, causing a significant impact on the magnets along the insertion regions (IRs), in particular the final focusing quadrupoles and the separation dipoles. Figure 10-1 also shows the breakdown of the debris components going through the aperture of the target absorber secondaries (TAS) absorber, a protection element consisting of a 1.8 m long copper core located 20 m from the IP and representing the interface between the detector and the accelerator. The TAS absorbers are only installed each side of the high luminosity detectors, ATLAS in P1, and CMS in P5, since their protection role, which is in fact limited to the first quadrupole, is not needed for luminosities up to $0.2 \times 10^{34}$ cm$^{-2}$ s$^{-1}$ [1].

Despite the fact that the number of particles per collision leaving the TAS aperture is more than one order of magnitude lower than the total number of debris particles, they carry about 80% of the total energy, implying that 40% of the released energy at the IP exits on each side of the experiments. At the nominal HL-LHC luminosity ($5 \times 10^{34}$ cm$^{-2}$ s$^{-1}$), this represents about 3800 W per side that is inevitably impacting upon the LHC elements and consequently dissipated in the machine and in the nearby equipment (e.g. electronics, racks, etc.) and in the tunnels walls.

It is fundamental to study how these particles are lost in order to implement the necessary protections for shielding sensitive parts of the LHC magnets and the machine. For these purposes, Monte Carlo simulations of particle interactions with matter play an essential role, relying on a detailed implementation of physics models and an accurate 3D description of the region of interest.

---

[*] Corresponding author: Francesco.Cerutti@cern.ch

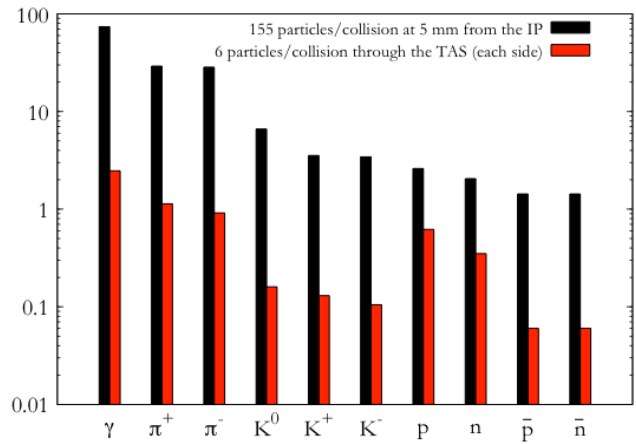

Figure 10-1: Number of debris particles per single proton–proton inelastic interaction at 5 mm from the interaction point (black) and at the exit of each 60 mm TAS aperture (red).

In addition to the luminosity debris, which dominates energy deposition in the vicinity of the collision points, regular and accidental beam losses represent other relevant sources of radiation. In particular, beam halo particles caught in the collimators (see Chapter 5) initiate hadronic and electromagnetic showers, mainly in the betatron and momentum cleaning IRs. The same happens with injection and dumping protection devices (see Chapter 14). Moreover, secondary particle showers are also originated by beam interactions with the residual gas inside the vacuum chamber the length of the accelerator, as well as with dust fragments falling into the beam path.

## 10.2 Power and dose evaluations concerning the triplet-D1 region

The LHC upgrade includes replacement of the inner triplet (IT) 70 mm Nb-Ti quadrupoles in P1 and P5 with the 150 mm coil aperture $Nb_3Sn$ quadrupoles, along with the new 150 mm coil aperture Nb-Ti dipole magnet and orbit correctors. Moreover, a corrector package (CP) that includes a skew quadrupole and eight high-order magnets (from sextupole to dodecapole, normal and skew, based on Nb-Ti technology) will be located between the triplet and D1.

As the first studies of radiation loads in the LHC upgrades have shown [2, 3], one could provide the operational stability and adequate lifetime of the IR superconducting magnets by using tungsten-based inner absorbers in the magnets. The goals are: i) reduce the peak power density in the inner $Nb_3Sn$ cable to below the quench limit with a safety margin; ii) keep the 3000 $fb^{-1}$ lifetime peak dose in the innermost layers of insulation and radiation loads on inorganic materials in the hottest spots of the coils below the known radiation damage limits; iii) keep the dynamic heat load to the cold mass at a manageable level.

### 10.2.1 FLUKA–MARS modeling

To design such a system in a consistent and confident way, coherent investigations have been undertaken with two independent Monte Carlo codes benchmarked in the TeV energy region and regularly used in such applications: FLUKA at CERN [4–7] and MARS15 (2014) at Fermilab [8–10]. The studies were done for 7 + 7 TeV p–p-collisions with a 295 µrad half-angle vertical crossing in IP1 (which had previously been found earlier to be the worst case) using DPMJET-III as the event generator.

An identical, very detailed geometry model was created and used in both codes with the same materials and magnetic field distributions in each of the components contained within the 80 m region from the IP through to the D1 dipole. An octagonal stainless steel beam screen, equipped with 6 mm tungsten absorbers on the mid-planes, is placed inside the cold bore along the triplet, the CP, and the D1, except in Q1, where the tungsten thickness is increased to 16 mm, compatible with the relaxed aperture requirements. The absorbers are between the beam screen and the 1.9 K beam pipe: they are supported by the beam screen, and thermally connected to it, whereas they have negligible contact with the cold mass. Therefore, from the point of view of energy deposition, the beam screen function is two-fold:

- it shields the coils from the debris by reducing the energy deposited;
- it removes a sizable part of the heat load from the 1.9 K cooling system, intercepting it at a higher temperature.

The present HL-LHC layout foresees six cryostats on each side of the IP: four for the triplet quadrupoles (Q1, Q2A, Q2B, and Q3), one for CP and one for D1. The distance between the magnets in the interconnections is 1.5–1.7 m, and an interruption of the beam screen is necessary therein. As a reasonable baseline, we assume here a 500 mm interruption of the tungsten absorbers in the middle of the interconnects. Figure 10-2 and Figure 10-3 show a 3D view of the model and details in the inner parts of the quadrupoles and orbit correctors.

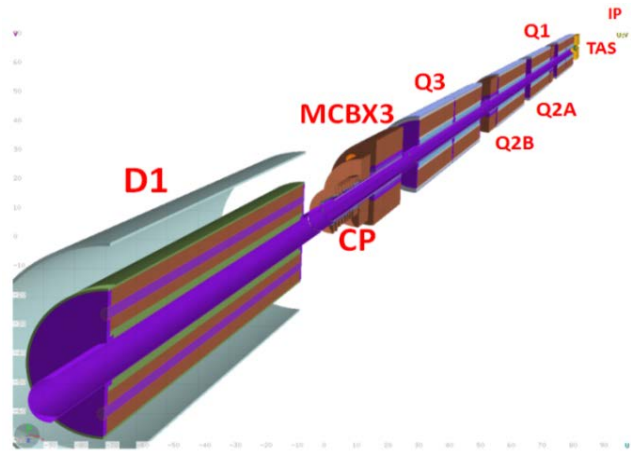

Figure 10-2: HL-LHC inner triplet with MCBX/CP correctors and D1 dipole

Fine-mesh distributions of power density as well as absorbed dose, neutron fluence, and displacements-per-atom (DPA), along with dynamic heat load in every IT component were calculated with FLUKA and MARS in high-statistics runs. The power density and dynamic heat load results are normalized to a luminosity of $5 \times 10^{34}$ cm$^{-2}$ s$^{-1}$, while all others are to 3000 fb$^{-1}$ integrated luminosity, corresponding to ~10–12 years of HL-LHC operation. Longitudinal scoring bins are 10 cm, and azimuthal ones are 2°. Radially, power density is averaged over the superconducting cable width, while dose, fluence, and DPA are scored within the innermost layer equal to 3 mm or its thickness, whichever is thinner.

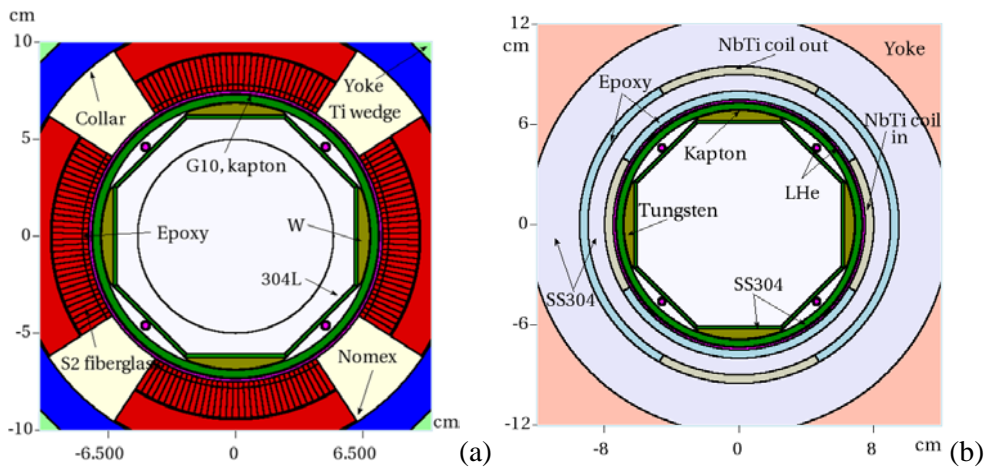

Figure 10-3: Details of the FLUKA–MARS model in the innermost regions of (a) the Q2–Q3 quadrupoles; (b) MCBX orbit correctors.

## 10.2.2 Operational radiation loads

Power density isocontours at the IP end of the cold mass of the Q2A quadrupole are shown in Figure 10-4(a). The longitudinal peak power density profile on the inner coils of the IT magnets at the azimuthal maxima is presented in Figure 10-4(b). Results from FLUKA and MARS are in excellent agreement, being observable discrepancies – which are naturally related to the use of fully independent tracking and scoring algorithms and physics models – largely within the safety margin to be recommended at the design stage. The peak value of 2 mW/cm$^3$ in the quadrupoles is 20 times less than the assumed quench limit of 40 mW/cm$^3$ in Nb$_3$Sn coils [2]. The peak value of ~1.5 mW/cm$^3$ in the Nb-Ti based coils of the correctors and D1 dipole is almost ten times less than the known quench limit of 13 mW/cm$^3$ in such coils [11].

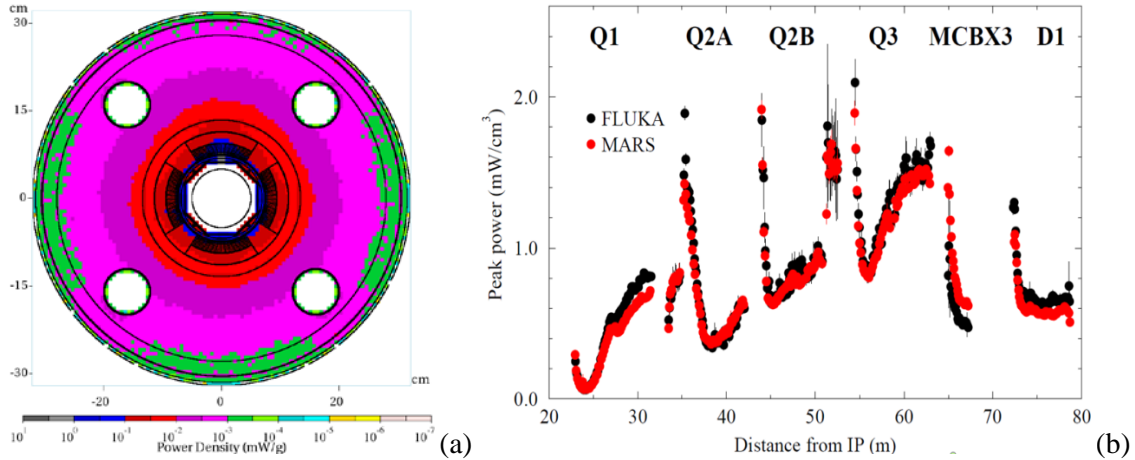

Figure 10-4: (a) Power density isocontours at the IP end of Q2A; (b) longitudinal peak power density profile on the inner coils of the IT magnets.

The total power dissipation in the inner triplet IT region from the IP1 collision debris splits roughly 50–50 between the cold mass and the beam screen with tungsten absorber, 630 W and 615 W, respectively, from the FLUKA calculations. MARS predicts values about 2% lower. For the 45 m effective length of the cold mass, the average dynamic heat load is ~14 W/m.

## 10.2.3 Lifetime radiation loads

The peak dose and DPA – the quantities that define radiation damage and the lifetime of the insulators and non-organic materials of the IT magnets, respectively – are calculated at the azimuthal maxima in the innermost tiny layers of each IT component shown in Figure 10-2.

The longitudinal peak dose profiles on the inner coils and insulating materials are presented in Figure 10-5(a). The values in the MCBX orbit correctors (located in the Q1–Q2A, Q2B–Q3, and Q3–D1 regions) are given for the epoxy layer (FLUKA) and kapton layer (MARS); see Figure 10-3 for details. Results from FLUKA and MARS are again in good agreement. The larger aperture IT magnets and the simulated implementation of tungsten absorbers perform very well, reducing the peak values of both power density and absorbed dose in the HL-LHC IR to the levels that correspond to LHC nominal luminosity.

The maximum peak dose in the coils is about 25 MGy for quadrupoles and ~15 MGy for the D1 dipole. The integrated peak dose in the IT magnet insulation reaches 30–36 MGy in the MCBX3 corrector, 28–30 MGy in the quadrupoles, and ~22 MGy in the D1 dipole. This is at or slightly above the common limits for kapton (25–35 MGy) and CTD-101K epoxy (25 MGy).

Degradation of the critical properties of inorganic materials in the IT magnets – Nb$_3$Sn and Nb-Ti superconductors, copper stabilizer, and mechanical structures – is usually characterized not by absorbed dose but by integrated neutron fluence and by DPA accumulated in the hottest spots over the magnet's expected lifetime. DPA is the most universal way to characterize the impact of irradiation on inorganic materials. In both FLUKA and MARS, all products of elastic and inelastic nuclear interactions as well as Coulomb elastic scattering (NIEL) of transported charged particles (hadrons, electrons, muons, and heavy ions) from ~1 keV to TeV energies contribute to DPA using Lindhard partition function and energy-dependent displacement efficiency. For neutrons at <20 MeV (FLUKA) and <150 MeV (MARS), the ENDF-VII database with NJOY99 processing is used in both the codes.

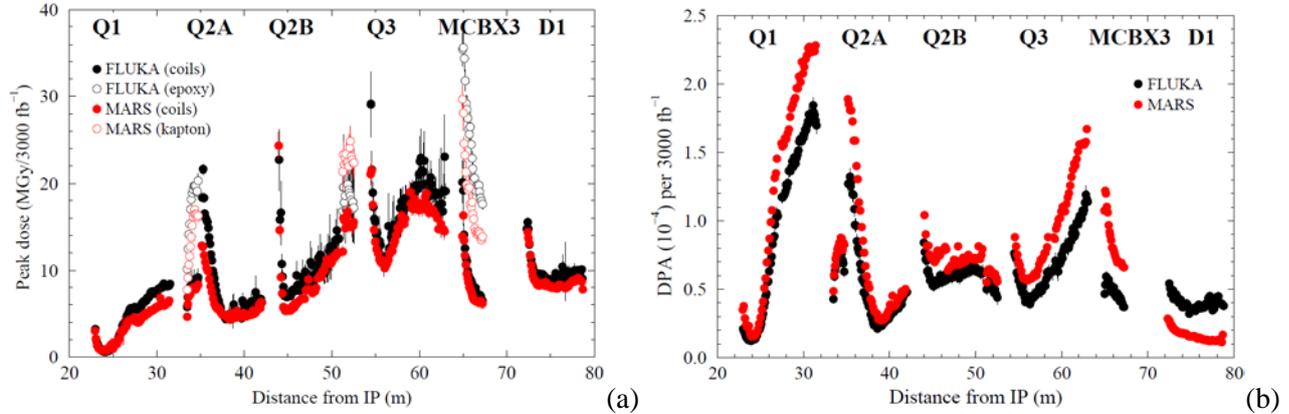

Figure 10-5: Longitudinal distributions of (a) peak dose on inner coils and nearby insulators; (b) peak DPA on inner coils.

The longitudinal peak DPA profiles on the IT magnet coils are presented in Figure 10-5(a). The peaks are generally observed at the inner coils; therefore, the data is given for these areas. With the vertical crossing in IP1, the MCBX3 orbit corrector is the exception with the peak in the outer coil in the vertical plane (see Figure 10-5). To see this effect, the MARS data in Figure 10-5(b) for MCBX3 is given for the outer coil, while FLUKA shows results for the inner coil as in all other magnets. Contrary to the power density and dose distributions driven by electromagnetic showers initiated by photons from $\pi^0$ decays, DPA peaks at the non-IP end of the Q1B quadrupole. At that location, about 70% of DPA is from neutrons with $E < 20$ MeV, ~25% from transported nuclear recoils above 0.25 keV/A, and the rest is due to other transported particles and non-transported recoils. The thicker Q1 shielding, while assuring a very effective dose reduction, plays a clear role in enhancing neutron production.

The peak in the Q1B inner coil is about $2 \times 10^{-4}$ DPA per 3000 fb$^{-1}$ integrated luminosity. In other IT components it is about $(1 \pm 0.5) \times 10^{-4}$. These numbers should be acceptable for the superconductors and copper stabilizer provided that there is periodic annealing during the collider shutdowns. Taking into account a good correlation of DPA with neutron fluence in the coils, one can also compare the latter with known limits. In the quadrupole coils, the peak fluence is ~$2 \times 10^{17}$ cm$^{-2}$ which is substantially lower than the $3 \times 10^{18}$ cm$^{-2}$ limit used for the Nb$_3$Sn superconductor. In the orbit corrector and D1 dipole coils, the peak fluence is ~$5 \times 10^{16}$ cm$^{-2}$ which is again lower than the $10^{18}$ cm$^{-2}$ limit used for the Nb-Ti superconductor. The integrated DPA in the magnet mechanical structures are 0.003 to 0.01 in the steel beam screen and tungsten absorber, ~$10^{-4}$ in the collar and yoke, and noticeably less outside. These are to be compared to a ~10 DPA limit for the mechanical properties of these materials. Neutron fluences in the IT mechanical structures range from $3 \times 10^{16}$ cm$^{-2}$ to $3 \times 10^{17}$ cm$^{-2}$, compared to the $10^{21}$ cm$^{-2}$ to $7 \times 10^{22}$ cm$^{-2}$ limits.

Peak dose on the beam screen was found to be of the order of several hundred MGy after 3000 fb$^{-1}$ integrated luminosity (up to 700 MGy in D1), mostly carried by electromagnetic particles. Its impact on carbon coating has to be considered.

## 10.3 Critical dependencies

The beam screen equipped with tungsten absorbers represents the backbone element for the protection of the IR magnets. Therefore, the details of its design play a crucial role in determining its actual effectiveness.

After the preliminary studies described in the previous section, new estimates were necessary to include:

- the real absorber material, Inermet 180, which has a density of 18 g cm$^{-3}$, about 8% less than pure tungsten, implying a reduced shielding performance;
- the first prototype taking into account the machinability of Inermet and the required size of the cooling tubes as dictated by preliminary cryogenics estimates;
- the reduction of the beam screen thickness (from 2 mm to 1 mm) necessary to let the structure respond elastically to possible deformations occurring during a quench.

Figure 10-6(a) shows a transverse section of the beam screen model (BS#2) embedding the abovementioned modifications. It can be compared to the model (BS#1) used in the calculations reported in the previous section (see Figure 10-3). The longitudinal peak dose profile on the inner coils of the IR magnets is presented in Figure 10-7 for BS#1 (black points) and BS#2 (red points). In the latter case, the accumulated peak dose turns out to be systematically higher all along the IR magnets, almost doubling its value in Q3 and reaching about 55 MGy in the MCBX3 corrector. Along Q2, most of the impacting debris, which is positively charged, is pushed by the magnetic field from the crossing angle side to the opposite side, i.e. from top to bottom in the assumed crossing scheme, where the outgoing beam points upwards. This moves the energy deposition peak through different azimuthal regions, which in the revised design (BS#2) are no longer shielded by the beam screen absorbers, hence yielding the resulting substantial increase. To fix this problem, we considered a third version of the beam screen (BS#3), where the Inermet absorbers were extended as much as possible to cover the coils towards the poles (see Figure 10-6(b)). The estimated peak dose distribution (green points in Figure 10-7(a) shows a significant improvement in the Q3–CP region, when compared to the BS#2 case. It should also be noticed that, mainly due to the reduced absorber density, the sharing of the total deposited power between cold mass and beam screen gets unbalanced, moving to 55–45 and making the heat released in the cold mass approach 700 W (at $5 \times 10^{34}$ cm$^{-2}$ s$^{-1}$).

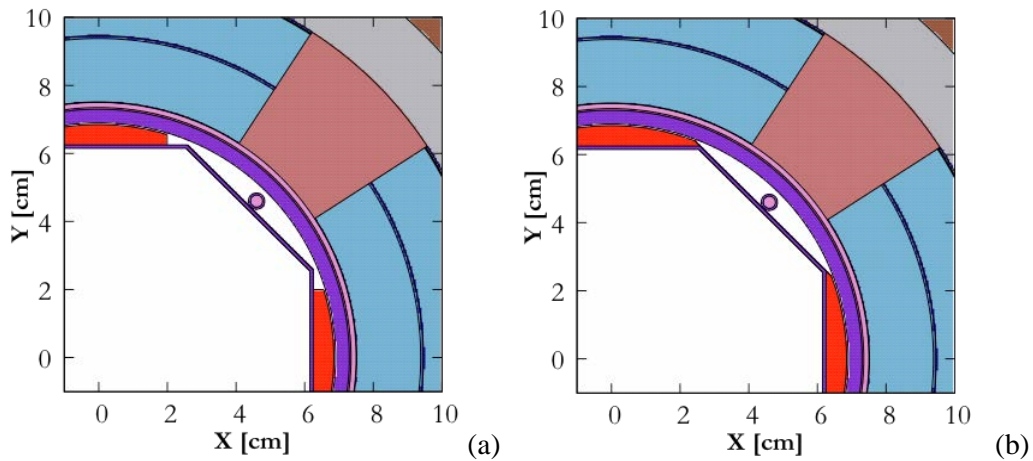

Figure 10-6: (a) Beam screen model as per the first conceptual design developed by WP12 (BS#2); (b) beam screen model with the modification of the absorbers driven by energy deposition considerations (BS#3).

Another crucial aspect is the longitudinal interruption of the beam screen and its absorbers, which is necessary between two consecutive cryostats in order to host a bellows and a BPM. As mentioned in the previous section, we initially assumed a 500 mm gap. Shorter gaps are possible if the BPMs are going to be equipped with absorber layers like those in the beam screen (see Chapter 12). To mimic this case, we looked at the effect of a 100 mm gap, which should be considered as the most optimistic case. The peak dose dependence on the gap length is presented in Figure 10-7(a) where the improvement achieved downstream of Q2A–Q2B, Q2B–Q3, and especially Q3–CP interconnects is visible, with a reduction from 55 MGy to 35 MGy in MCBX3 for the BS#2 design.

Therefore, the actual implementation of the absorber layers in the design of both the beam screen and the relevant BPMs considerably affects the maximum dose expected in the IT coils.

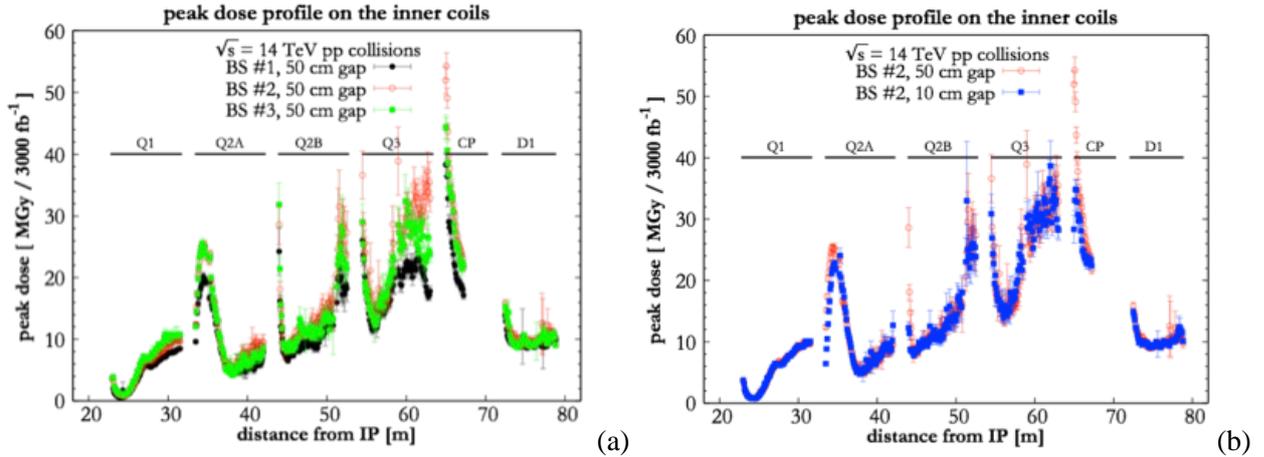

Figure 10-7: Longitudinal distributions of peak dose on the inner coils of the IR magnets referring to (a) different beam screen designs; (b) different lengths of the beam screen gap in the interconnects.

### 10.4 Impact on the matching section and protection strategy

The HL upgrade foresees changes in the IR1 and IR5 matching section, implying a new target absorber neutral (TAN) absorber with larger apertures, a new recombination dipole D2 and Q4 (both with larger twin apertures, together with their respective correctors). The present baseline also foresees the Q5 replacement with the present Q4 (providing a larger aperture) and the installation of crab cavities. From Q6 onwards, no hardware modifications are expected inside the IRs.

With respect to the present LHC, the variation of three ingredients mainly impacts the energy deposition in the matching section elements. They are:

- the distance between the separation/recombination dipoles (reduced from 86 m to 67 m);
- the aperture of the upstream magnets (IT quadrupoles, CP correctors, and D1);
- the TAN design, which has to comply with larger beam separation and sizes.

As a consequence, the number of debris particles entering the matching section per primary collision is much larger than in the case of the current machine. This is illustrated in Figure 10-8, where the debris particle distributions at the exit of the TAN outgoing beam pipe are shown for the LHC (Figure 10-8(a)) and the HL-LHC (Figure 10-8(b)). The number of protons is increased by about 30% (from 0.12 to 0.16 protons/collision), while the number of photons and neutrons is about seven times higher (from 0.06 to 0.41 particles/collision). Note that in the case of the HL-LHC optics the beam size at this location is about twice as large as that of the LHC optics. Therefore, a collimator set at the same aperture in beam sigmas turns out to be less effective in intercepting debris particles, as clearly revealed in the figure by the number of particles left inside the beam envelope.

In the present LHC machine, a network of target collimator long physics debris (TCL) collimators secures the protection of the cold magnets in the matching section. There is one copper TCL in front of D2 (TCL4, installed during LS1), one copper TCL in front of Q5 (TCL5, the only one already present and operating during Run 1), and one tungsten TCL in front of Q6 (TCL6, whose installation during LS1 was triggered by specific Roman pot operation scenarios). However, at nominal luminosity not all TCLs are necessary to keep the heat load below the quench level. It was shown that the single TCL4 set at 15 $\sigma$ is sufficient to maintain the peak power density below 0.3 mW/cm$^3$ in all matching section magnets [12].

Conversely, in the case of the HL-LHC, all of the TCLs are indispensable for magnet protection. For the purposes of the estimates presented here, the jaws of all the collimators are assumed to be made of tungsten (Inermet 180) because of its greater absorption efficiency (higher density and atomic number than copper). Moreover, additional fixed target collimator long mask (TCLM) masks are required in order to further shield the magnet aperture. The masks are supposed to be placed between the magnet cryostat and the TCL collimator (where present). Since the hottest spot is typically located on the IP side of each magnet, a shielding strategy based on a beam screen equipped with thick absorbers, as in the IT region, does not pay here. On the other hand, the masks, which have the same shape and aperture as the beam screen of the protected magnet and a radial thickness sufficient to shadow the downstream coils, can ensure an effective interception of the shower coming from the upstream elements. Sensitivity to mask presence and length has been investigated, and also aiming to comply with integration issues that are particularly challenging in the TAN–D2 region.

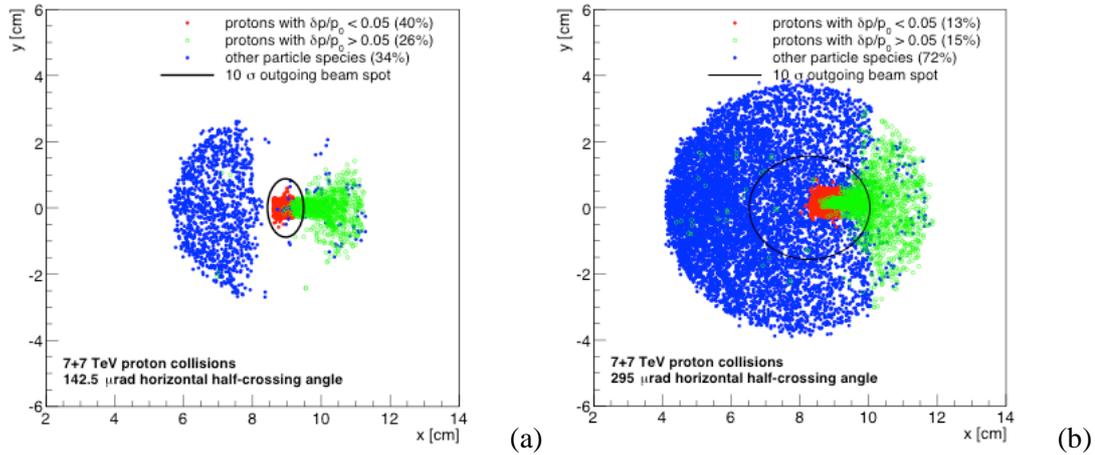

Figure 10-8: Debris particle distribution at the exit of the outgoing beam pipe of the TAN, (a) for the LHC; (b) for the HL-LHC. Red points indicate protons with magnetic rigidity within 5% with respect to beam protons and green points indicate protons with lower magnetic rigidity. Blue points indicate neutral particles (photons and neutrons). In both cases the same number of collision is simulated. The black ellipse shows the 10 $\sigma$ outgoing beam spot.

Preliminary longitudinal profiles of the peak power density along D2, Q4, Q5, and Q6 are shown in Figure 10-9. All of the TCLs are set to 10 $\sigma$, but for TCL4 a 20 $\sigma$ aperture is also explored. Dedicated masks turn out to be necessary for Q5 and Q6, while their need is less obvious for D2 and Q4, for which further studies are being carried out taking into account refined specifications concerning the TAN and TCL4 aperture and position. Considering lifetime issues, the installation of 50 cm masks in front of Q5 and Q6 keeps the respective peak dose values within 20 MGy after 3000 fb$^{-1}$.

The heat load due to collision debris on the most exposed crab cavity (the first one on the outgoing beam) was estimated to be about 0.2 W. The maximum power density is located on the cavity internal plate and is about 0.4 mW/cm$^3$. It corresponds to ~3 MGy after the target HL-LHC integrated luminosity. The contribution from beam–gas interactions is expected to be 10 times smaller, for a conservative residual gas density of 10$^{15}$ H$_2$-equivalent molecules/m$^3$.

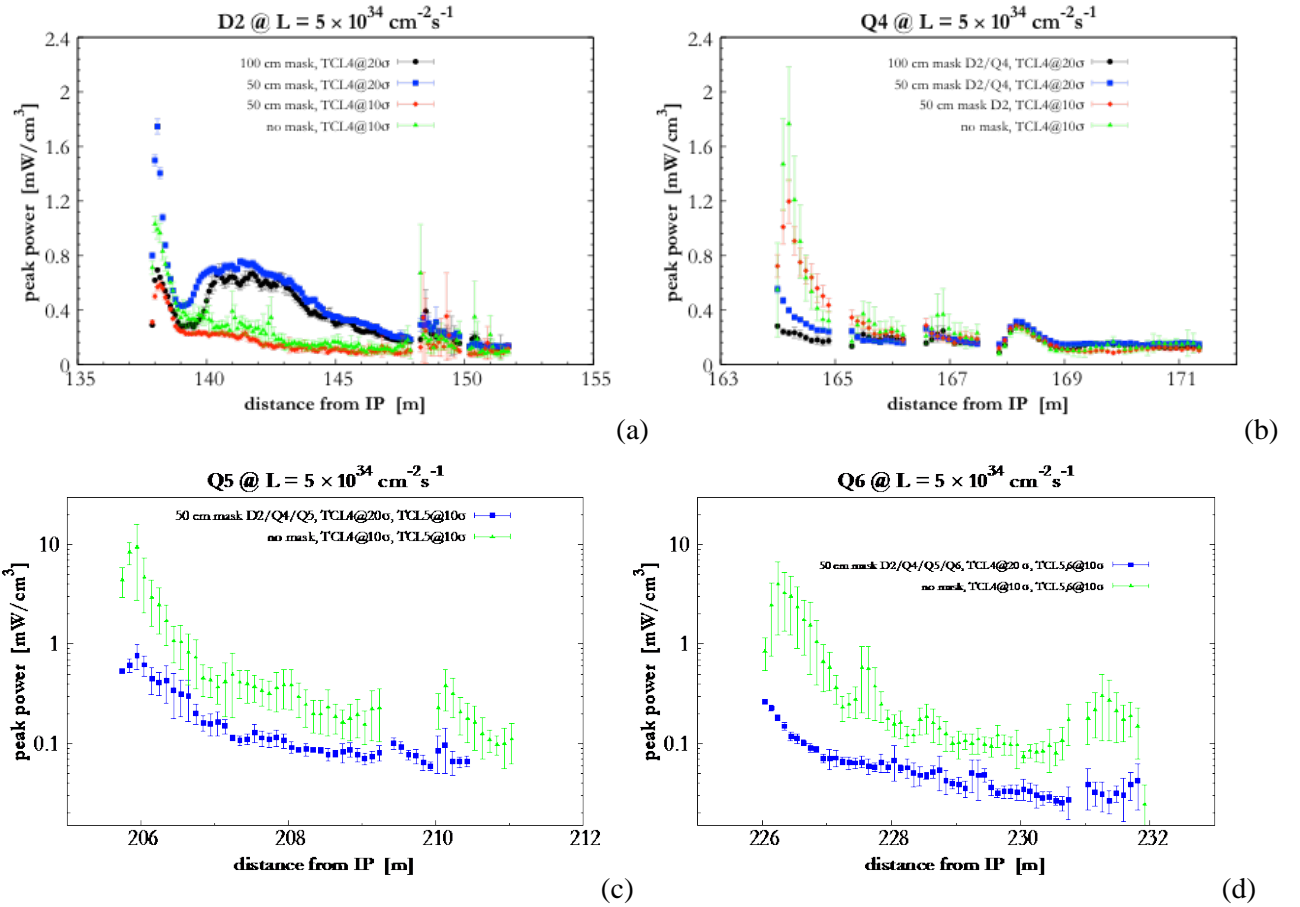

Figure 10-9: Longitudinal profile of peak power density along (a) D2; (b) Q4; (c) Q5, log scale; (d) Q6, log scale, at the HL-LHC target luminosity of $5 \times 10^{34}$ cm$^{-2}$ s$^{-1}$. Different layouts and settings are compared. Error bars indicate statistical uncertainties.

## 10.5 Exposure of the superconducting links

Cold powering of the HL-LHC magnets foresees moving the power converters to the surface [13]. The consequences of this new configuration are:

- safer long-term operation of powering equipment (power converters, current leads, and associated auxiliary devices), being located in a radiation-free environment;
- safer access of personnel to equipment for maintenance, repair, and diagnostic and routine test interventions;
- reduced time of interventions on power converters and current leads;
- more free space in the beam areas, which becomes available for other equipment.

The connecting lines will be made of MgB$_2$. The link cold mass contains SC cables that are connected at one end, in the tunnel, to the Nb-Ti magnet busbar operating in liquid helium; and the other end to the bottom end of the current leads maintained at a maximum temperature ($T_{CL}$) of about 20 K in a helium gas environment, see Figure 10-10 [14].

The radiation impact on the MgB$_2$ SC links is evaluated from different points of view. Boron consumption by thermal neutrons is not a concern. The total number of neutrons escaping from the first quadrupole of the new triplet is of the order of $10^{21}$, over an integrated luminosity of 3000 fb$^{-1}$. Considering the amount of $^{10}$B in the links, consumption can be estimated to be much less than 0.01%.

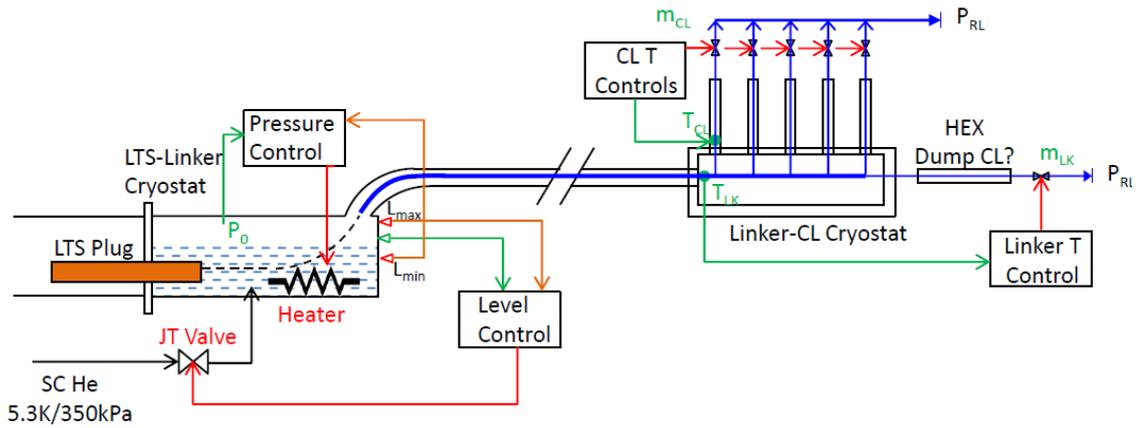

Figure 10-10: Cooling scheme of the cold powering system. Lefthand side: low temperature (LT) side, where there are the connections with the SC magnet cable. Righthand side: connection with the current leads (CL).

The links in P1 and P5 consist of a multi-cable assembly as shown in Figure 10-11(a). Its model as implemented in the FLUKA geometry is also shown in Figure 10-11(c).

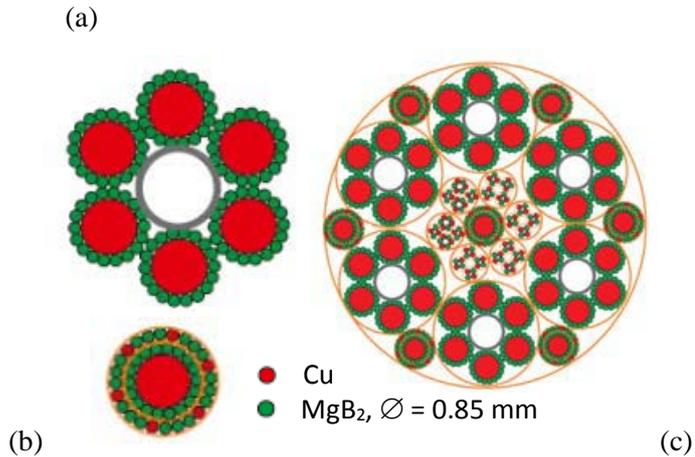

Figure 10-11: (a) 20 kA $MgB_2$ cable, $\varnothing = 19.5$ mm, two $MgB_2$ concentric cables for 3 kA total current, $\varnothing = 8.2$ mm. (b) 150 kA cable assembly for LHC P1 and P5 (6 × 20 kA, 7 × 3 kA, 4 × 0.4 kA, 20 × 0.12 kA), $\varnothing = 65$ mm. (c)The FLUKA model of the cable.

The calculated dose maps for the horizontal and vertical routing of the SC link inside the shuffling module in P1 are shown in Figure 10-12. The obtained values (up to about 1 MGy) are not expected to affect the link operation (provided that the chosen insulator is radiation resistant). The maximum DPA induced in the links is of the order of $10^{-6}$. Preliminary simulations for P5 indicate the same outcome as for P1.

The modelling of the cable in P7 is in progress; so, as a first approximation, the same cable as in P1 and P5 has been used in order to evaluate its level of exposure to radiation. In this case, the latter is originated by beam losses in the collimators, from which energetic particle showers develop. The link routing has been implemented at about 1 m above the beam. The hottest region is downstream of the primary collimators, close to the passive absorber that shields the warm dipoles. Assuming that the ratio between yearly losses in the collimators and accumulated luminosity does not get significantly worse than what was achieved with the present machine, the resulting maximum dose and DPA are again not of concern for the Link operation and long-term integrity.

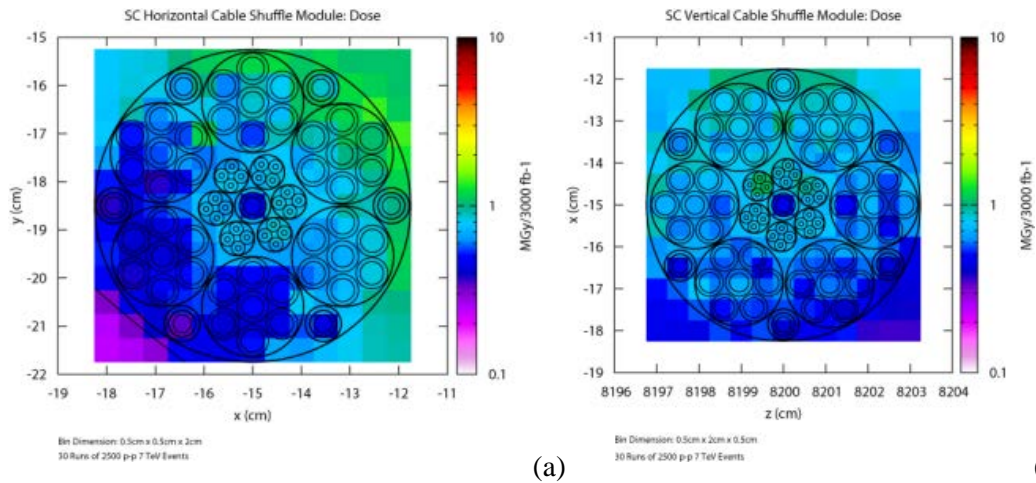

Figure 10-12: Maps of dose released in the SC Link in the P1 shuffling module. (a) Superconducting horizontal cable shuffle module dose; (b) Superconducting vertical cable shuffle module dose. Values (in MGy) are normalized to 3000 fb$^{-1}$.

## 10.6 Radiation To Electronics

A specific problem is represented by the electronics' sensitivity to radiation. The above described particle debris emerging from the IP, together with the additional loss contribution from beam-gas interactions, will impact equipment present in the areas adjacent to the LHC tunnel (UJs, RRs). Installed (present or future) control systems are either fully commercial or based on COTS components, both possibly affected by radiation. This includes the immediate risk of SEE and a possible direct impact on beam operation, as well as in the long-term cumulative dose effects (impacting the component/system lifetime) that additionally have to be considered.

For the tunnel equipment in the existing LHC, radiation was only partially taken into account as a design criteria prior to construction, and most of the equipment placed in adjacent and partly shielded areas was not conceived nor tested for their actual radiation environment. Therefore, given the large amount of electronics being installed in these areas, during the past years a CERN-wide project called Radiation To Electronics (R2E) [15] has been initiated to quantify the danger of radiation-induced failures and to mitigate the risk for nominal beams and beyond to below one failure per week. The respective mitigation process included a detailed analysis of the radiation fields involved, intensities and related Monte Carlo calculations; radiation monitoring and benchmarking; the behaviour of commercial equipment/systems and their use in the LHC radiation fields; as well as radiation tests with dedicated test areas and facilities [15, 16].

In parallel, radiation-induced failures were analyzed in detail in order to confirm early predictions of failure rates, as well as to study the effectiveness of implemented mitigation measures. Figure 10-13 shows the actual number of SEE failures measured during 2011 and 2012 operations, the achieved improvement (please note that the failure rate measured during 2011 already benefitted from mitigation measures implemented during 2009 and 2010), as well as the goal for operation after LS1 and during the HL-LHC era.

Aiming for annual luminosities of up to 300 fb$^{-1}$, it is clear that machine availability has to be maximized during the HL-LHC period in order to successfully achieve the physics goal. This implies that existing electronics control systems are either installed in fully safe areas, sufficiently protected by shielding, or are made adequately radiation tolerant.

The last statement implies that existing equipment, as well as any future equipment that may be installed in R2E critical areas, must be conceived in a specific way.

Radiation damage to electronics is often considered in space applications. However, it is important to note that the radiation environment encountered at the LHC, the high number of electronics systems and components partly exposed to radiation, as well as the actual impact of radiation-induced failures, differ strongly from the context of space applications. While for the latter application design, test, and monitoring standards are already well defined, additional constraints, but in some cases also simplifications, have to be considered for the accelerator environment.

The mixed particle type and energy field encountered in the relevant LHC areas is composed of charged and neutral hadrons (protons, pions, kaons, and neutrons), photons, electrons, and muons ranging from thermal energies up to the GeV range. This complex field has been extensively simulated by the FLUKA Monte Carlo code and benchmarked in detail for radiation damage issues at the LHC [17, 18]. As discussed above, the observed radiation is due to particles generated by proton–proton (or ion–ion) collisions in the LHC experimental areas, distributed beam losses (protons, ions) around the machine, and to beam interacting with the residual gas inside the beam pipe. The proportion of the different particle species in the field depends on the distance and on the angle with respect to the original loss point, as well as on the amount (if any) of installed shielding material. In this environment, electronics components and systems exposed to a mixed radiation field will experience three different types of radiation damage: displacement damage, damage from the TID, and SEEs. The latter range from single or multiple bit upsets (SEUs or MBUs), transients (SETs) up to possible destructive latch-ups (SELs), destructive gate ruptures, or burn-outs (SEGRs and SEBs).

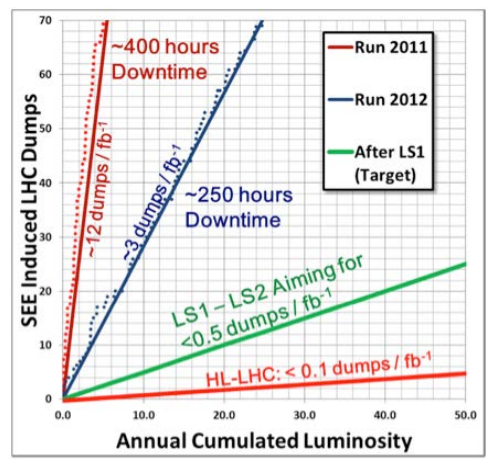

Figure 10-13: LHC beam dumps due to single-event effects against beam luminosity. Dots (2011 and 2012) refer to measurements, whereas lines show annual averages for both past and future operations.

The first two groups are of cumulative nature and are measured through TID and non-ionizing energy deposition (non-ionizing energy losses (NIEL), generally quantified through accumulated 1 MeV neutron equivalent fluence), where the steady accumulation of defects causes measurable effects that can ultimately lead to device failure. As for stochastic SEE failures, they form an entirely different group, since they are due to direct ionization by a single particle, and are able to deposit sufficient energy through ionization processes to perturb the operation of the device. They can only be characterized in terms of their probability of occurring as a function of accumulated high energy (>5–20 MeV) hadron (HEH) fluence. The probability of failure will strongly depend on the device as well as on the flux and nature of the particles. In the context of the HL-LHC, several tunnel areas close to the LHC tunnel, and partly not sufficiently shielded, are or are supposed to be equipped with commercial or not specifically designed electronics that are mostly affected by the risk of SEEs, whereas electronics installed in the LHC tunnel will also suffer from accumulated damage in the long term [19].

Three distinct areas are to be distinguished.
- The accelerator between the inner triplet and Q6: in this area no active electronics is installed and any future installation should clearly be avoided as radiation levels clearly exceed the usability domain of commercial components and also pose significant constraints to custom designs, including application specific integrated circuits (ASICs).

- The adjacent (shielded) areas: where the shielding has been maximized before and during LS1 and where radiation levels become acceptable for characterized and qualified custom electronics. The equipment and selected electronics components have to be qualified for both SEEs and TID. For the first, it is important that radiation tests are carried out in a representative radiation environment, or adequate safety margins have to be added. Fully commercial systems (COTS based) are still to be avoided in these areas. In case they are required, their failure impact and respective mitigation has to be studied in the context of accelerator operation.
- The dispersion suppressor area: given the fact that the magnets have to be protected against quenches, it is likely that the leakage into this area must not increase significantly with respect to nominal LHC operation. A detailed quantification is, however, needed to coherently design the required control electronics, again both for SEE effects as well as for their lifetime (TID) to comply with the stringent availability requirement, as shown in Figure 10-13.

During the first years of LHC operation, the radiation levels in the LHC tunnel and in the (partly) shielded areas have been measured using the CERN RadMon system [20], which is dedicated to the analysis of radiation levels possibly impacting installed electronics equipment. Table 10-1 summarizes the level of accumulated HEH fluence measured during 2012 for the most critical LHC areas where electronics equipment is installed and that are relevant for the HL-LHC project, together with the expected annual radiation levels for nominal LHC performance (50 fb$^{-1}$/y). The HEH fluence measurements are based on the RadMon reading of the SEUs of SRAM memories whose sensitivity has been extensively calibrated in various facilities [21]. The results obtained during 2012 LHC proton operation show that the measurements compare very well with previously performed FLUKA calculations, and that observed differences can actually be attributed to changes of operational parameters not considered in the calculations [22]. In a first approximation, the measured radiation levels can also be used to extrapolate towards the HL-LHC by purely scaling with annual luminosity (see the last two columns of Table 10-1); however, keeping in mind that operational and layout parameters (beam energy, crossing angle, TAN design, absorbers, etc.) can have a non-negligible impact on the final values, especially for the RRs close to the matching section. The resulting values clearly indicate that any equipment installed in the LHC tunnel will not only suffer SEE failures, but will also be heavily impacted by TID effects, thus limiting the equipment's lifetime.

Table 10-1: Predicted and measured annual HEH fluence in critical shielded areas for a cumulated ATLAS/CMS luminosity of 15 fb$^{-1}$ during 2012 operations, then extrapolated based on the measurement to the expected nominal and HL-LHC performance (50 fb$^{-1}$/y for nominal and 300 fb$^{-1}$/y for HL-LHC performance, except for P8 where 2012 can already be considered as almost nominal and HL-LHC refers to a five-fold increase). For the HL-LHC an estimate for corresponding annual TID levels is also given.

| LHC area | Prediction (HEH/cm$^2$) | Measured (HEH/cm$^2$) | Nominal (HEH/cm$^2$) | HL-LHC (HEH/cm$^2$) | HL-LHC (Dose/Gy) |
|---|---|---|---|---|---|
| UJ14/16 | $1.4 \times 10^8$ | $1.6 \times 10^8$ | $5 \times 10^8$ | $3 \times 10^9$ | 6 |
| RR13/17 | $2.0 \times 10^8$ | $2.5 \times 10^8$ | $8 \times 10^8$ | $5 \times 10^9$ | 10 |
| UJ56 | $1.6 \times 10^8$ | $1.5 \times 10^8$ | $5 \times 10^8$ | $3 \times 10^9$ | 6 |
| RR53/57 | $2.0 \times 10^8$ | $2.5 \times 10^8$ | $8 \times 10^8$ | $5 \times 10^9$ | 10 |
| UJ76 | $2.1 \times 10^7$ | $6.0 \times 10^7$ | $2 \times 10^8$ | $1 \times 10^9$ | 2 |
| RR73/77 | $2.9 \times 10^7$ | $5.0 \times 10^7$ | $2 \times 10^8$ | $1 \times 10^9$ | 2 |
| UX85B | $4.3 \times 10^8$ | $3.5 \times 10^8$ | $4 \times 10^8$ | $2 \times 10^9$ | 4 |
| US85 | $1.3 \times 10^8$ | $8.8 \times 10^7$ | $9 \times 10^7$ | $4 \times 10^8$ | 1 |

A first calculation aimed to assess the radiation levels in the UJ/UL/UP areas close the new IT during HL-LHC operation was carried out, implementing the dedicated shielding presently in place. Figure 10-14 shows the expected annual HEH fluence, which confirms the order of magnitude of the respective extrapolation reported in the first row of Table 10-1.

Concerning the radiation in the RR area adjacent to the matching section, HL-LHC simulations are at this stage still premature due to the unknowns regarding optics and layout. Nevertheless, specific studies were performed for the present LHC, evaluating the contribution of the relevant source terms, which in this case include both the collision debris and the beam interactions with the residual gas, as shown in Figure 10-15. A significant effect in increasing the fluence levels – to be re-evaluated for the HL-LHC machine – is played here by the TCL6 collimator, due to its position.

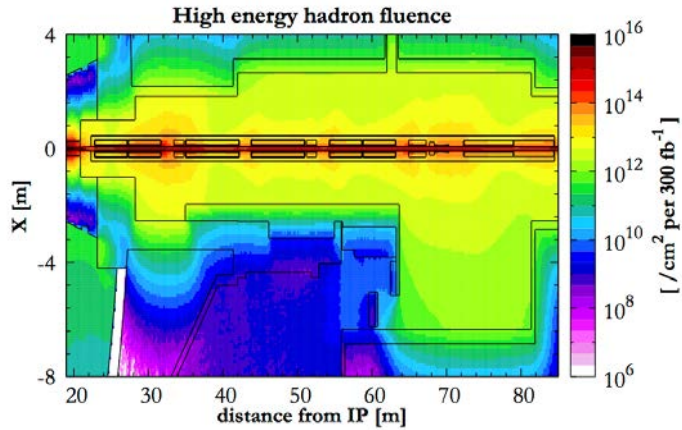

Figure 10-14: Annual HEH fluence expected in the IT region and in the adjacent UJ/UL/UP areas at P1 during the HL-LHC era (normalized to 300 fb$^{-1}$).

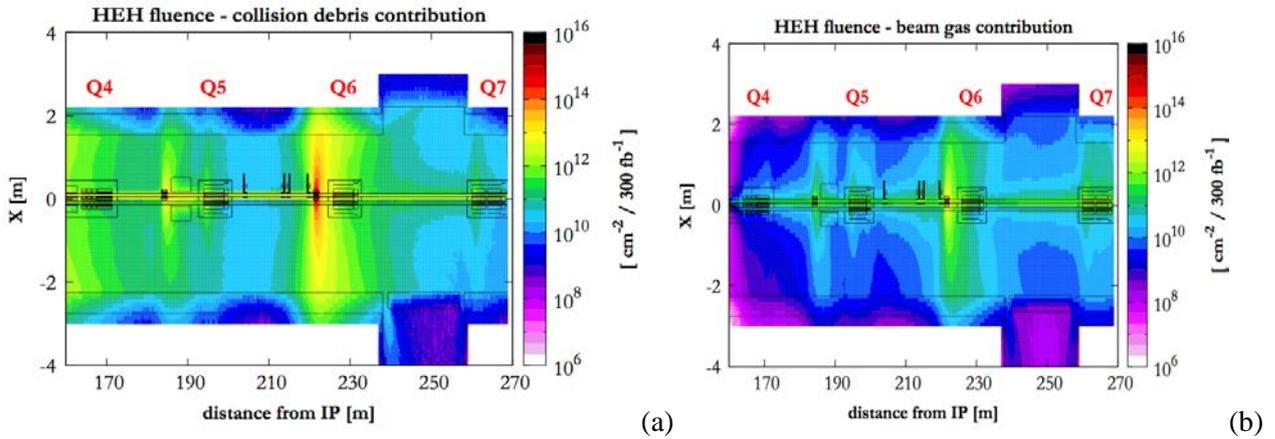

Figure 10-15: Cumulated HEH fluence expected in the matching section region and in the adjacent RR area at P5 for the operation of the present LHC at 7 TeV beam energy. (a) Contribution from the collision debris normalized to 300 fb$^{-1}$. (b) Contribution from interactions of the outgoing beam (Beam 1) with the residual gas, normalized to nominal current over $3 \times 10^7$ s (i.e. the time needed to accumulate 300 fb$^{-1}$ at nominal luminosity) and to a gas density of $10^{15}$ H$_2$-equivalent molecules/m$^3$. TCL collimator settings: 15 $\sigma$ (TCL4), 35 $\sigma$ (TCL5), 10 $\sigma$ (TCL6).

Any control equipment (commercial or based on commercial components) to be installed in these areas clearly has to be proven to be sufficiently radiation tolerant. For comparison, as mentioned earlier, during the last years of operation we have already observed a number of radiation-induced failures of commercial equipment at radiation levels corresponding to $10^8$–$10^9$ cm$^{-2}$/y (which is about 1 000–10 000 times more than what one would get at the surface due to cosmic radiation).

The analysis of the performed radiation tests, as well as the experience acquired during LHC Run 1 operation allowed the deduction of an acceptable limit of $10^7$ cm$^{-2}$ y$^{-1}$ annual radiation level, leading to the definition of so-called protected areas (in terms of overall risk of radiation-induced failures). Therefore, for the HL-LHC any installation of non-tested (and not specifically designed) electronics equipment in the UJs, part of the ULs, and RRs is clearly to be avoided or subjected to a detailed analysis process before an exceptional installation can be granted under the following conditions:

- the equipment is not linked to any safety system;
- the failure of the equipment will not lead to a beam dump;
- the failure of the equipment does not require quick access (thus lead to downtime);
- there is no any other operational impact (loss of important data, etc.).

In all other cases requiring installation in critical areas, a respective radiation-tolerant electronics development must be considered from a very early stage onward. Related expertise exists at CERN within the equipment groups, the R2E project, and a dedicated working group [23].

In a first approximation and limiting the total number of exposed systems, the above-mentioned annual radiation design level of $10^7$ cm$^{-2}$ y$^{-1}$ can also be chosen as acceptable, aiming to achieve an overall performance of less than one radiation-induced failure per one or two weeks of HL-LHC operation.

For operation critical equipment, the HL-LHC project already foresees radiation-tolerant developments at an early stage of the design phase, taking into account that:

- for the LHC-tunnel: in addition to SEEs cumulative damage also has to be considered for both existing and future equipment;
- for partly shielded areas (UJs, RRs, ULs): cumulative damage should be carefully analyzed but can most likely be mitigated by preventive maintenance (detailed monitoring mandatory); however, radiation-tolerant design is mandatory in order to limit SEE-induced failures;
- the knowledge of radiation-induced failures and radiation-tolerant development within the equipment groups and in the overall accelerator and technology sector has to be maintained and further strengthened;
- the access and availability of radiation test facilities (both inside CERN and outside) has to be ensured, providing efficient support to equipment groups;
- building on the experience obtained during the LHC R2E project and in view of the HL-LHC time-scale, it is important that the expertise of and support to radiation-tolerant developments (currently available through the Radiation Working Group [23]) are maintained and ensured from the early project stages onwards.